\documentclass[journal=jacsat,manuscript=article]{achemso}

\usepackage{chemformula} 
\usepackage[T1]{fontenc} 
\usepackage{siunitx}
\usepackage{xcolor}

\makeatletter
\newcommand\thefontsize{The current font size is: \f@size pt}
\makeatother

\newcommand{\vm}[1]{\textcolor{blue}{#1}}

\usepackage{soul}

\author{David H\"ahnel}
\author{Jens F\"orstner}
\email{jens.foerstner@uni-paderborn.de}
\author{Viktor Myroshnychenko}
\email{viktor.myroshnychenko@gmail.com}
\affiliation[Paderborn University] {Paderborn University, Theoretical Electrical Engineering, Warburger Str. 100, 33098 Paderborn, Germany}
\title[Efficient nonlinear wavefront shaping by dielectric metasurfaces]{Efficient nonlinear wavefront shaping by dielectric metasurfaces}


\begin{document}

\begin{tocentry}

\includegraphics[scale=1]{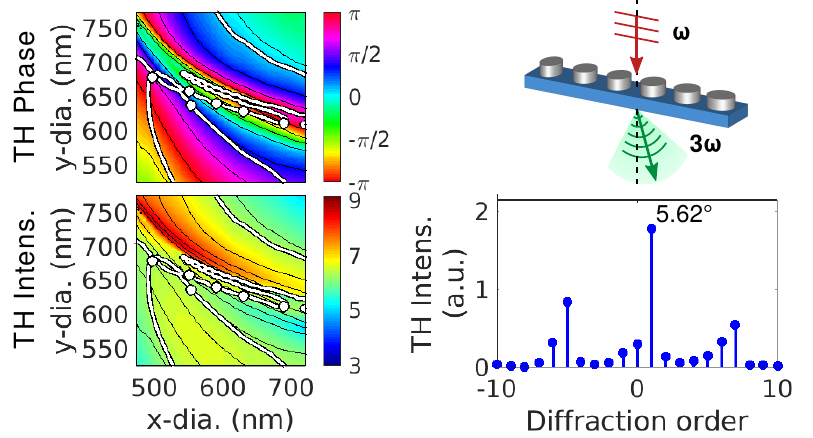}




\end{tocentry}

\begin{abstract}
Dielectric metasurfaces provide a unique platform for efficient harmonic generation and optical wavefront manipulation at the nanoscale. Tailoring phase and amplitude of a nonlinearly generated wave with a high emission efficiency using resonance-based metasurfaces is a challenging task which often requires state-of-the-art numerical methods. Here, we propose a simple yet effective approach combining a sampling method with a Monte Carlo approach to design the third-harmonic wavefront generated by all-dielectric metasurfaces composed of elliptical silicon nanodisks. Using this approach, we demonstrate the full nonlinear $2\pi$ phase control with a uniform and highest possible amplitude allowing us to design metasurfaces operating as third harmonic beam deflectors capable of steering light into a desired direction with high emission efficiency. The high amplification of the third-harmonic intensity emitted at the first diffraction order by a factor up to $\sim 500$ is achieved as compared to the best results reported for silicon-based metasurfaces so far. We anticipate that the proposed approach will be widely applied as alternative to commonly used optimization algorithms with higher complexity and implementation effort for the design of metasurfaces with other holographic functionalities.
\end{abstract}

\section{Introduction}
Artificially designed metasurfaces comprising a spatial arrangement of planar arrays of subwavelength resonators have recently attracted tremendous scientific and technological attention as they exhibit a number of extraordinary properties including the ability for complex optical wavefront engineering \cite{Chen2018,Zhang2021,Paniagua2020,Pertsch2020,Solntsev2021}. These ultrathin structures permit to control and manipulate the amplitude, phase, direction, and polarization state of the incident light along the surface at the subwavelength scale, opening a new avenue within the dynamic field of nanophotonics and nanooptics. This makes them promising candidates in a wide range of exciting devices ranging from polarization converters \cite{Yu2012,Abouelatta2021,Kruk2016} and optical sensors \cite{Guo2014,Guo2019} to flat lenses\cite{Aieta2012,Aieta2015} and complex holograms \cite{Almeida2016,Huang2013,Wang2016,Balthasar2017}.

The wavefront control is commonly achieved by using geometric-phase (Pancharatnam-Berry phase) metasurfaces \cite{Balthasar2017,Ahmed2020,Frese2021,Liu2020} or resonant (Huygens) metasurfaces \cite{Staude2013,Aoni2019,Kepic2021,Wang2018}. In the former case, the wavefront is modulated by locally positioning and rotating the resonators in a certain way on the surface, whereas in the latter case it is realized by varying size and shape of the individual resonators which, in turn, control the excitation of a variety of optical resonances locally hosted within them. Hence, Huygens metasurfaces usually exhibit higher conversion efficiency and can operate under the excitation with linearly polarized light. In this context, metasurfaces made of high-index dielectric resonators have gained great popularity in recent years as an alternative to their metallic counterparts, as they possess crucial advantages and offer more degree of freedom in wavefront design. First, the dielectric resonators exhibit very low dissipation loss and thus have high laser-damage threshold. Second, they are capable to host electric and magnetic multipolar resonances which can be spectrally tailored, to overlap and even swap positions in the spectrum, while in common metallic nanoparticles only electric resonances are fully supported \cite{Myroshnychenko2008} due to their weak magnetic response within the optical range \cite{AsenjoGarcia2012}. Furthermore, the richness of these resonances can lead to the excitation of nonradiative modes and ultra sharp Fano resonances in these structures. Finally, in contrast to metallic nanoparticles, whose localized surface plasmon resonances are confined at the metal-dielectric interface \cite{Myroshnychenko2008}, the Mie modes of high-index dielectric resonators are accommodated inside the whole volume, increasing the excitable, optical mode volume \cite{Krasnok2018}. All these advantages make dielectric metasurfaces appealing candidates for the efficient nonlinear light generation \cite{Li2017,SChen2018,Grinblat2021}. 
A variety of exciting functionalities in nonlinear domain have recently been shown including nonlinear imaging\cite{Gao2018,Schlickriede2020,Kruk2022}, beam steering \cite{Liu2020,Wang2018}, and ultrafast optical modulation \cite{Maier2020}. In particular, the designed nonlinear metasurfaces made of silicon, germanium, or aluminium gallium arsenide nanoparticles have demonstrated large enhancement of the conversion efficiency and possibility for advanced nonlinear wavefront control\cite{Celebrano2015,Huttunen2019,Shcherbakov2014,Grinblat2017}.

Many existing papers are aimed at either an amplitude amplification or phase control of the harmonic signal, however, there is a lack of publications on full nonlinear wavefront control, which is a challenging task, especially, when considering the tailoring of the continuous phase in a certain range at constant and preferentially maximal amplitude using resonance-based metasurfaces. These challenges arise mainly from the narrow-band behavior of the high-order electric and magnetic resonances and couplings between them which are strongly influenced by the geometrical parameters of the resonators \cite{Wang2018}. This implies that small changes in the geometrical parameters of the resonators can result in large conversion efficiency jumps and large phase variations, making such metasurfaces sensitive to fabrication imperfections. The problem is compounded by the fact that generated nonlinear radiation depends strongly on variations of the phase and amplitude of the fields not only at pump frequency but also at high harmonic frequency. Furthermore, to achieve above mentioned functionalities, the resonators are usually packed in arrays whose collective responses have to be taken into account as well. All of the above mentioned factors require sophisticated optimization for precise determination of the proper set of resonators. For this purpose, different methods have been extensively applied, such as gradient descent method \cite{Wang2018,Mansouree2019}, topology optimization \cite{Deng2016,Sell2017,Phan2019}, evolutionary optimization technique \cite{Huntington2014,Wiecha2017} in conjunction with electro-magnetic solvers like finite-difference time-domain method, finite integration technique, etc. Besides, different machine learning techniques and deep neural network-based approaches have been increasingly considered recently to further accelerate the metasurface design process \cite{Malkiel2018,Liu2018,Kudyshev2020}.

In this work, we propose a simple and robust sampling method in conjunction with Monte Carlo (MC) simulation to design and optimize a nonlinear wavefront of Huygens all-dielectric metasurfaces. We demonstrate our approach for a classical metasurface composed of elliptical nanodisk resonators made of silicon, that generate a third-harmonic (TH) radiation under external illumination, placed on a silicon dioxide substrate. Information about the nonlinear phase and amplitude are locally encoded through adjustment of lateral dimensions of nanodisks. A MC simulation is employed to explore a geometrical parameter space and find optima with with enhanced conversion efficiency. The sampling method is then applied to select a collection of resonators generating a TH field with a phase discretely spanning the full $2\pi$ range at the uniform and highest possible amplitude. We exemplary show the power of our approach by designing and optimizing dielectric metasurfaces which act as nonlinear beam deflectors. The high amplification of TH generation emitted along the prescribed angle is achieved. Our combined approach clearly demonstrates the possibility to design and optimize nonlinear Huygens metasurfaces with a variety of functionalities without usage of sophisticated optimization methods.

\newpage
\section{Results and discussion}
\subsection{Model of metasurface for nonlinear beam deflection}
\begin{figure}
\includegraphics[scale=1]{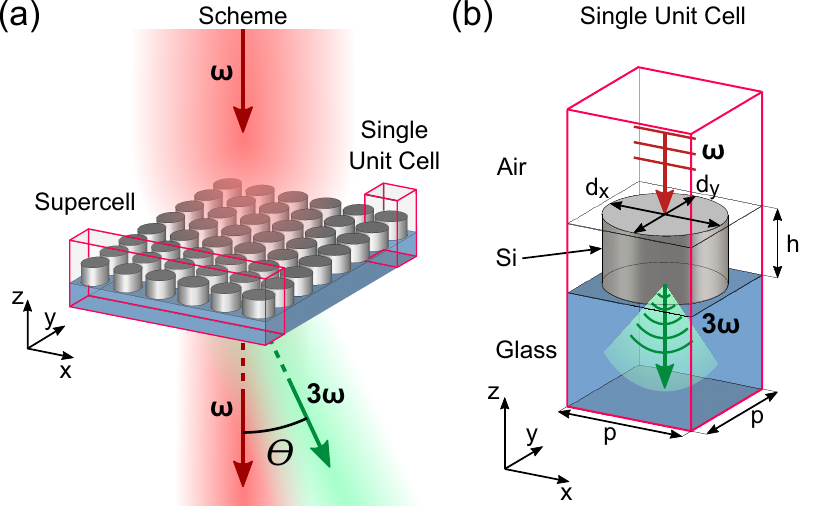}
\caption{(a) Schematic view of a dielectric metasurface for nonlinear beam deflection. The metasurface is composed of an array of supercells each of which consists of $N$ silicon resonators in a form of elliptical cylinders with different lateral dimensions placed on a SiO$_2$ substrate (blue block). The metasurface is illuminated from the top by linear $x$-polarized light at the fundamental frequency $\omega$ (red arrow) and deflects the TH wave $3\,\omega$ (green arrow) into the substrate by the angle $\theta$. The incident and generated TH waves experience a $2\pi$ phase shift over the length of each supercell. (b) A model of a periodic square unit cell used in the design and optimization. Phase and amplitude control of the TH field is achieved by tailoring the resonance condition of the Si resonator through variation of the cell periodicity $p$, resonator height $h$ and elliptical base diameters $d_x$ and $d_y$.} 
\label{fig:fig1}
\end{figure}

To implement our approach, we utilize a metasurface composed of a set of silicon elliptical cylinder disks placed on a silicon dioxide substrate \cite{Wang2018}. It exhibits TH generation under external illumination since silicon belongs to the group of centrosymmetric materials with a significant third-order nonlinearity. We aim to design a metasurface which produces a TH beam deflection into a desired direction (metadeflector), as illustrated in Figure~\ref{fig:fig1}a. The deflection of the TH output wave is caused by a stepwise linear phase shift along the surface induced by the dielectric resonators. Such a metasurface is composed of a periodic array of supercells each of which consists of a set of resonators emitting nonlinear fields with a linear phase profile spanning the full $0-2\pi$ range. Additionally, we pursue maximization of the nonlinear emission efficiency with a uniform amplitude for all resonators. The TH steering angle $\theta$ for such a supercell can be calculated as $\theta=\sin^{-1}(\lambda_{\mathrm{TH}}/Np)$, where $\lambda_{\mathrm{TH}}$ is the TH wavelength in the output medium, $N$ is the number of resonators in the supercell, and $p$ is the dimension of the single unit cell. The periodic properties of such metasurfaces can be exploited to reduce the model in the simulations to one supercell, i.e. a single line of resonators schematically depicted in Figure~\ref{fig:fig1}a, with application of periodic boundary conditions. The overall response of such a supercell is assumed to depend only on the properties of the individual resonators. Therefore, in order to minimize memory space and time consumption required for the design and optimization process, in a first step we can further reduce the simulation model to a unit cell with a single resonator from the entire structure, as shown in Figure~\ref{fig:fig1}b. In this case, the main assumption exploited is that the coupling between resonators of different shape is comparable to equal shapes thus not affecting the TH far-field much. The phase and amplitude of the nonlinear beam can be controlled by varying the geometrical parameters such as the square unit cell size $p$, the resonator height $h$ and elliptical base diameters $d_x,d_y$. Calculations of the linear and nonlinear electromagnetic fields were performed by employing the finite element method. Details of the numerical simulations are provided in the Methods section. Finally, we note that this work is mainly focused on the design of the metasurfaces with a full-phase control of the high efficiency harmonic field and does not explore the origin and nature of electromagnetic resonances inducing TH emission.

\subsection{Sampling method for design of TH phase and amplitude}
\begin{figure*}[t]
\includegraphics[scale=1]{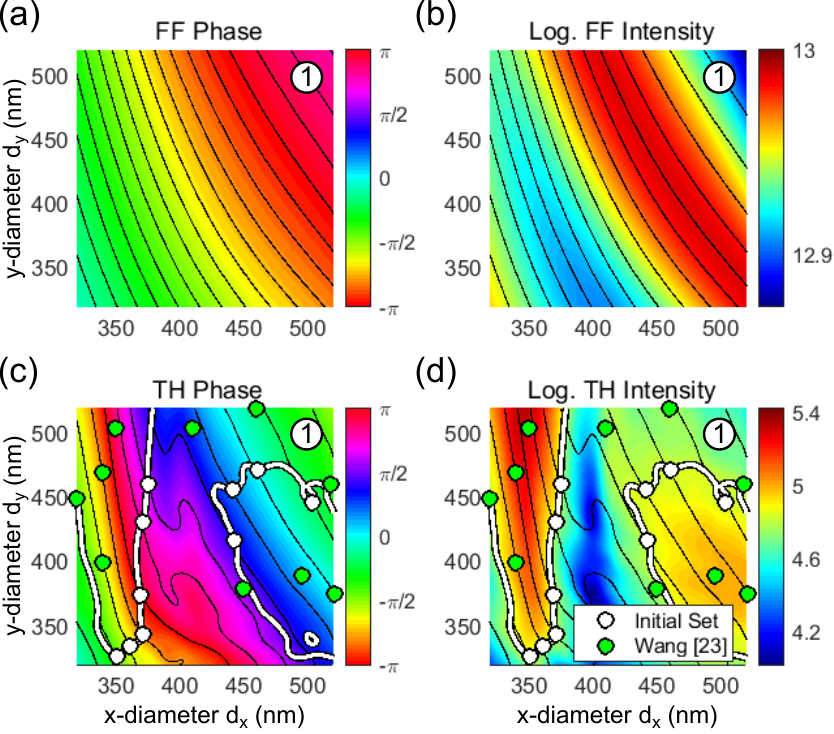}
\caption{Sampling method for the design of nonlinear metasurfaces with full nonlinear phase and amplitude control. (a) Phase and (b) intensity maps of the FF electric field at $\lambda=$\SI[mode=text]{1615}{\nano\metre} in transmission direction as a function of nanodisk elliptical diameters $d_x$ and $d_y$. The periodicity of the unit cell and the nanodisk height are fixed at $p$=\SI[mode=text]{550}{\nano\metre} and $h$=\SI[mode=text]{617}{\nano\metre}, respectively. Black curves correspond to the contour isolines of the phase angles along which the function has constant values. (c) Phase and (d) intensity maps of the TH electric field at $\lambda=$\SI[mode=text]{538.3}{\nano\metre} with contour isolines. The values of the phase contour isolines (black curves) are equidistantly distributed in $2\pi/10$ steps. The white curves in (c,d) represent the contour isolines for the highest intensity value that continuously covers the full phase angle range of $2\pi$ in the phase map (c). Ten nanodisks with elliptical diameters $d_x$ and $d_y$ marked by the white dots are selected using a sampling method based on the detection of a predefined number of intersections between phase and intensity contours. The green dots indicate the nanodisks optimized by a gradient descent method in the work of Wang et al. \cite{Wang2018}.}
\label{fig:fig2}
\end{figure*}

The work of Wang et al. \cite{Wang2018} was used as a starting point providing us with a preoptimized set of parameters for the nonlinear silicon metadeflector. For the design of our initial metasurface, we also use a pump wavelength of \SI[mode=text]{1615}{\nano\metre}, a lattice periodicity of \SI[mode=text]{550}{\nano\metre}, a nanodisk height of \SI[mode=text]{617}{\nano\metre}, and elliptical base diameters in the range from \SI[mode=text]{320}{\nano\metre} to \SI[mode=text]{535}{\nano\metre}. To achieve the desired nonlinear beam steering, the phase and amplitude information for each nanodisk in the supercell at the TH wavelength (\SI[mode=text]{538.3}{\nano\metre}) is required. For this purpose, we computed maps which relate a given TH phase and amplitude to the values of elliptical basis diameters $d_x$ and $d_y$, as illustrated in Figure~\ref{fig:fig2}c,d \cite{Gigli2021}. Additionally, the contour isolines of the TH phase angles along which the function has constant values (black curves) are added. The values of the phase contours are equidistantly distributed in the $0-2\pi$ range determining also the number of distinct resonator geometries. The phase map calculated in this parameter range clearly covers a full $2\pi$ relative phase shift. Taking inspiration from a sampling method proposed by \textit{Almeida et al.} for design of phase-only plasmonic holograms in the nonlinear domain \cite{Almeida2016}, we select a set of ten resonators required to construct a similar metadeflector as in Ref. \cite{Wang2018}. This is done by detecting a predefined number of intersections (white dots) between phase contours (black curves) and amplitude contour at the highest possible uniform amplitude (white curve) which covers the full phase angle range of $0-2\pi$, as shown in the maps of Figure~\ref{fig:fig2}c,d. This selection procedure makes the method very simple, adaptive, and robust to any number of resonators and any phase and amplitude profiles required. Interestingly, in contrast to the supercell optimized by the gradient descent method in the work of Wang et al. \cite{Wang2018}, demonstrating varying harmonic amplitudes among the selected nanodisks (green dots), the supercell designed by the sampling method has a perfectly uniform amplitude among the nanodisks and, importantly, a similar magnitude as the average amplitude from \cite{Wang2018}. The phase and amplitude maps at FF together with contour isolines of the phases are also shown for comparison in Figure~\ref{fig:fig2}a and b, respectively. The smooth change of the linear phase and amplitude suggests detuning from the resonance peaks of nanodisks in this parameter range. The large deviations between the phase and amplitude maps at FF and TH confirm the impact of Mie resonances at both pump and harmonic frequencies on the phases and amplitudes of the radiated nonlinear waves. 

\subsection{Monte Carlo simulation for TH amplitude enhancement}
\begin{figure}[t]
\includegraphics[scale=1]{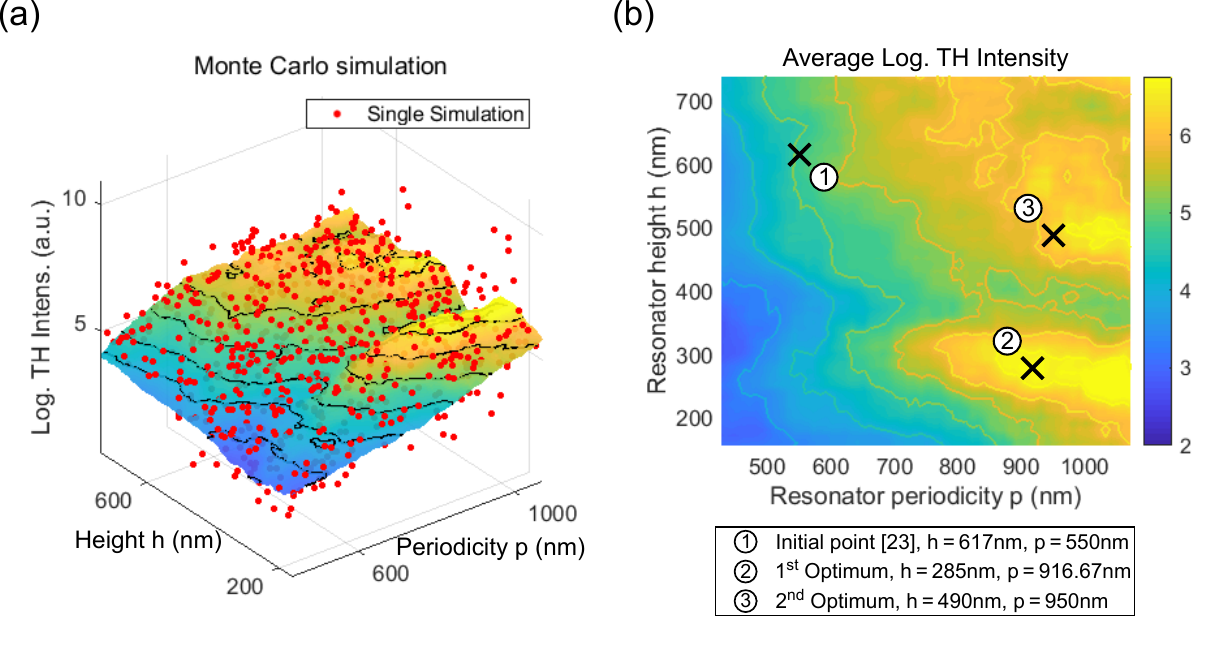}.
\caption{Monte Carlo simulation for design/optimization of metasurfaces with enhanced TH generation. (a) Scatter plot of TH intensity in transmission direction (red dots) calculated as a function of four geometrical parameters, the lattice periodicity, nanodisk height, and the two elliptical base diameters. The values of these parameters are randomly generated in the predefined parameter ranges for a large collection of samples. The surface plot represents the averaged TH intensity (log scale) calculated by applying a 2D moving average procedure to the scattered data and plotted as a function of the lattice periodicity and nanodisk height. (b) Projection of the surface plot in (a) on the plane, showing regions of high intensity. Three black crosses in the map indicate parameter sets for the lattice periodicity and nanodisk height selected for design of metadeflectors: initial set 1 demonstrated in Figure~\ref{fig:fig2} ($p=$\SI{550}{\nano\metre}, $h=$\SI{617}{\nano\metre}), optimum set 2 ($p=$\SI{917}{\nano\metre}, $h=$\SI{285}{\nano\metre}), and optimum set 3 ($p=$\SI{950}{\nano\metre}, $h=$\SI{490}{\nano\metre}).}
\label{fig:fig3}
\end{figure}

 Aiming to further increase the intensity of the TH radiation, we employ a Monte Carlo simulation which permits us to explore the full parameter space and find promising geometrical parameters furnishing the enhanced TH signal. This optimization approach in combination with the sampling method introduces an exploration dynamics into the design process which is typically not addressed by standard optimization algorithms. The procedure consists of the following steps. First, we pick out four geometric parameters affecting the TH intensity, namely the lattice periodicity, the nanodisk height, and two elliptical base diameters. In the spirit of MC simulations, the values of these parameters are randomly generated in predefined parameter ranges and the TH intensity of this single ``random'' resonator in the periodic unit cell is calculated as explained in the Methods section. The large collection of such samples $\sim 24000$ was generated in order to accurately detect regions of high intensity in a four-dimensional (4D) parameter space. Figure~\ref{fig:fig3}a illustrates the resulting scatter plot of TH intensity (red dots) as a function of the lattice periodicity and nanodisk height, by projection disregarding the actual values of elliptical base diameters. Note that only a small part of dense scattered data are plotted for better visibility. Next, these intermediate results are processed by applying a 2D moving average procedure to the scattered data points and calculating the averaged TH intensity as a function of the lattice periodicity and resonator height. The ``moving'' window has a size of \SI[mode=text]{45}{\nano\metre}$\times$\SI[mode=text]{45}{\nano\metre} and is shifted with a step size of \SI[mode=text]{10}{\nano\metre}. The resulting surface plot of average TH intensity presented in Figure~\ref{fig:fig3}a unveils tendencies in the parameter space. The projection of the surface plot on the plane represented in Figure~\ref{fig:fig3}b indicates several distinguishable maxima with considerable high intensities. Finally, the lattice periodicity and resonator height which resulted in highest average TH intensity can be selected in this map, reducing the parameter set to two nanodisk diameters used for final manipulation of the TH phase and amplitude. The highest maximum in the map, marked by a black cross and named as optimum set 2, corresponds roughly to the lattice periodicity of \SI{917}{\nano\metre} and nanodisk height of \SI{285}{\nano\metre}. We also select another prominent maximum, optimum set 3, with the periodicity of \SI{950}{\nano\metre} and nanodisk height of \SI{490}{\nano\metre}. The other maxima located at the higher values of parameters are neglected in order to consider fabrication restrictions and to avoid higher order diffraction. The initial geometry optimized by Wang et al. \cite{Wang2018} and explored in Figure~\ref{fig:fig2} ($p=$\SI{550}{\nano\metre}, $h=$\SI{617}{\nano\metre}) is also marked in the map as initial set 1. Importantly, the comparison of initial and optimum sets indicates the increase of average TH intensity by approximately two orders. Note that this is solely the average amplification of the radiated TH intensity from the homogeneous metasurface.

\subsection{Design of MC-optimized metadeflectors}
\begin{figure*}
\includegraphics[scale=1]{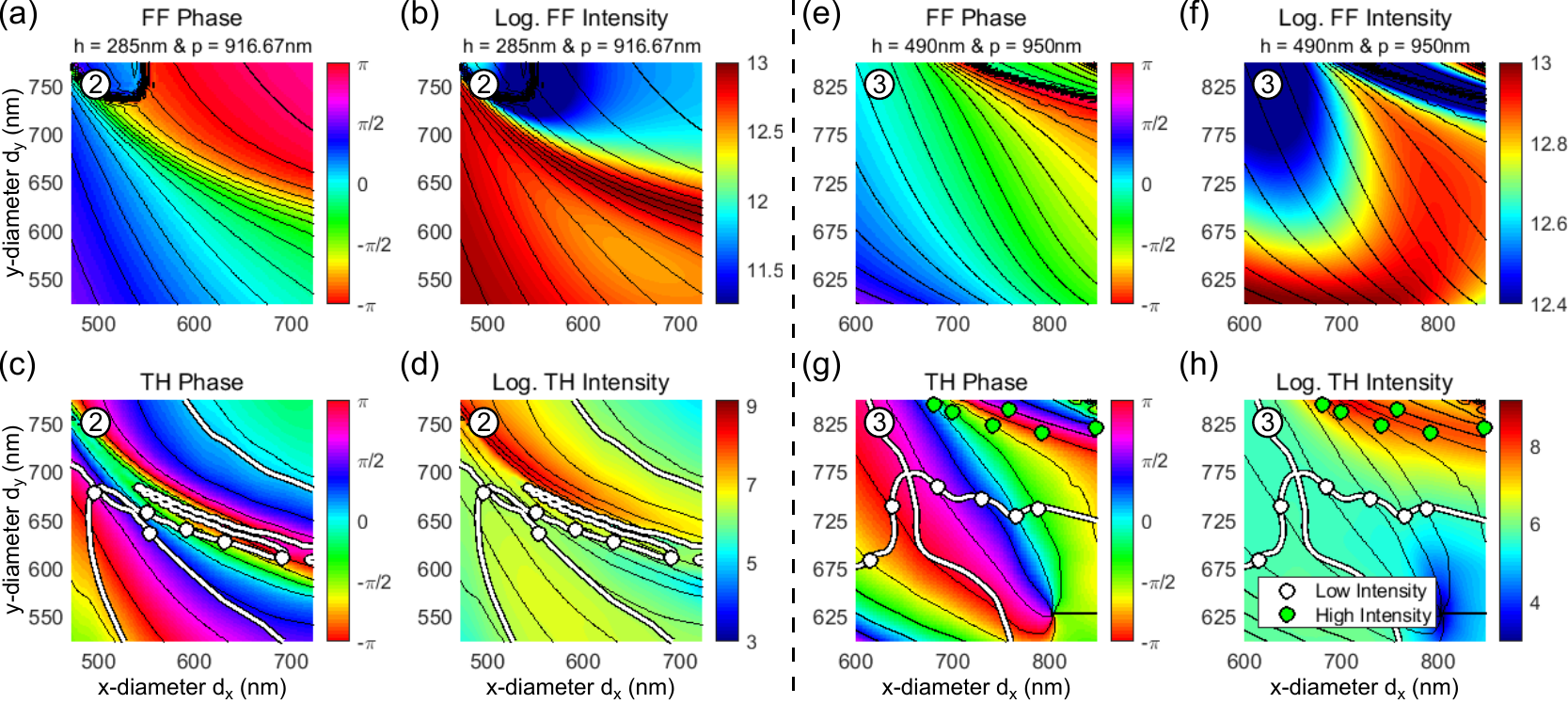}
\caption{Sampling method for design of the nonlinear metasurfaces with the MC-optimized amplitude. (a-d) Phase and intensity maps of the electric field in transmission direction at (a,b) FF and (c,d) TH for optimum set 2 with $p$=\SI[mode=text]{916.67}{\nano\metre} and $h$=\SI[mode=text]{285}{\nano\metre} optimized by a MC procedure in Figure~\ref{fig:fig3}b. Black curves in maps correspond to the equidistantly distributed contour isolines of the phase angles along which the function has constant values. White curves in (c,d) represent the contour for the highest TH intensity value that covers the full $2\pi$ phase range in the phase map (c). Six individual nanodisks marked by the white dots are selected using a sampling method. (e-h) Same as in (a-d) for optimum set 3 corresponding to the periodicity $p$=\SI{950}{\nano\metre} and the nanodisk height $h$=\SI{490}{\nano\metre}. Two sets of nanodisks are selected using a sampling method (white dots) and manually in a region of a significantly higher intensity (green dots).}
\label{fig:fig4}
\end{figure*}

Having two sets of optimal parameters for the lattice periodicity and nanodisk height gained from the MC optimization procedure (optimum sets 2 and 3 in Figure~\ref{fig:fig3}), we computed the corresponding {linear and} TH phase and intensity maps by sweeping across the elliptical diameters of the nanodisk, as presented in Figure~\ref{fig:fig4}. For each case, the sampling method presented above was applied to select a set of resonators (white dots) spanning the phase of $0-2\pi$ range with equal spacing and the highest possible uniform amplitude (white curves). Noting the significant differences in the values of the initial and optimum lattice periodicities and aiming at a nearly same deflection angle for the TH beam, we selected six nanodisks featuring approximately the same supercell length as initial set 1 with ten resonators explored in Figure~\ref{fig:fig2}. A close look at the TH phase map for optimum set 2 shown in Figure~\ref{fig:fig4}c reveals an abrupt change of the phase along the winding amplitude contour (white curves) caused mainly by the overlap of the excited optical resonances. This implies a small variation in some selected nanodisk diameters (along $y$-axis) and, thus, requires high fabrication accuracy. In contrast, for optimum set 3 (Figure~\ref{fig:fig4}g,h), the phase changes smoothly along the amplitude contour (white curves) resulting in a high difference in the selected nanodisks diameters, but the TH intensity decreases by one order of magnitude compared to the one for optimum set 2. Pursuing further the high TH radiation, we focus our attention on a resonant area of high intensity in the upper right corner of Figure~\ref{fig:fig4}h. This TH enhancement is caused by a strong resonance feature at FF manifested in Figure~\ref{fig:fig4}f (blue area in the upper-right corner). There is no perfectly uniform amplitude contour spanning the $2\pi$ phase range in this region. However, by relaxing the uniform amplitude condition, the set of six required resonators possessing a nearly uniform TH intensity with $2\pi$ phase coverage can be manually selected, as marked by green dots. It is named optimum set 3 of high intensity. In summary, the TH intensity amplification by a factor $\sim 30$ (optimum set 2), $\sim 7$ (optimum set 3, low intensity), and $\sim 1000$ (optimum set 3, high intensity) with respect to the reference intensity (initial set 1) is achieved by employing MC simulation for optimization.

\subsection{TH beam steering}
\begin{figure}[t]
\includegraphics[scale=1]{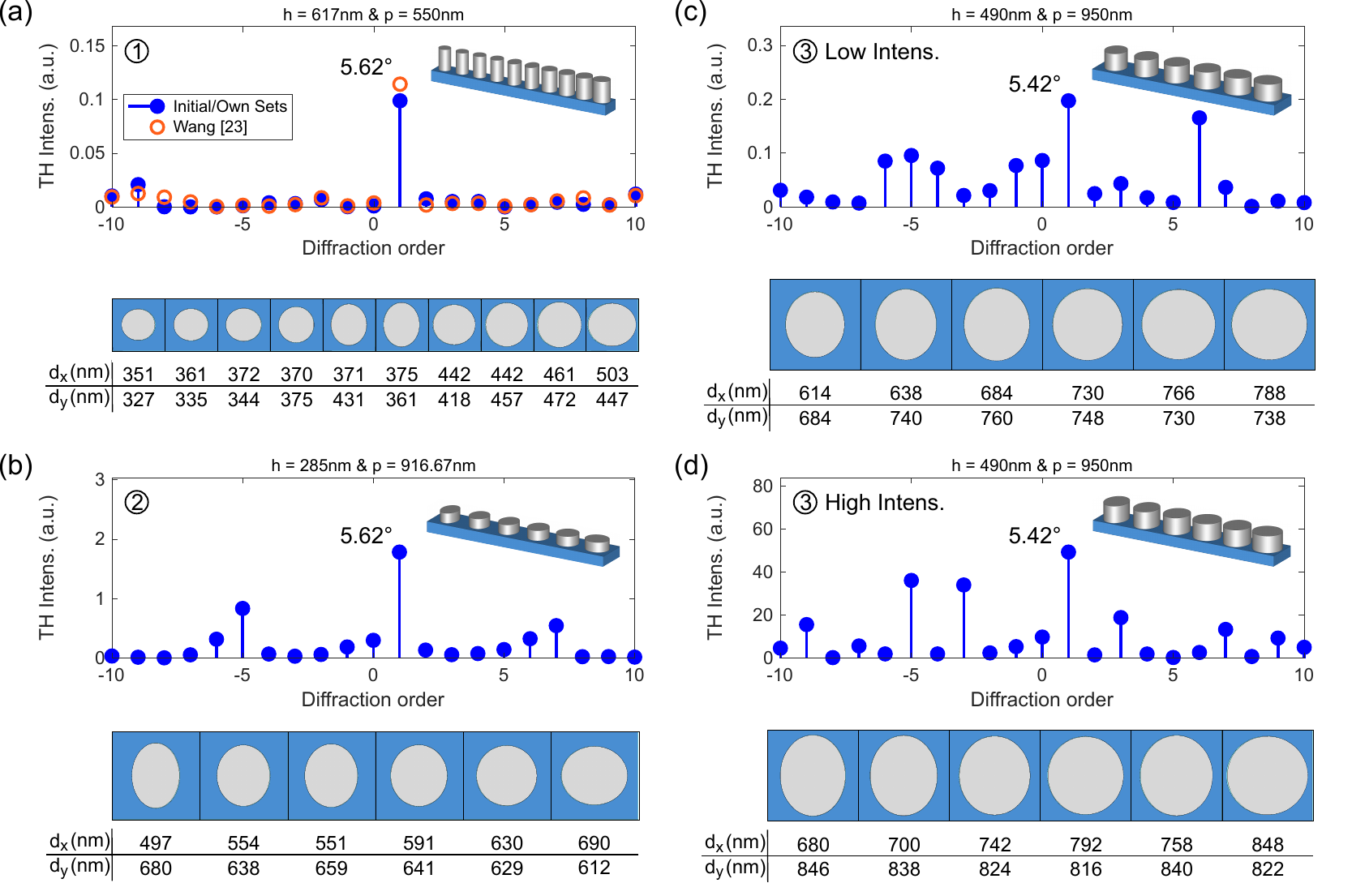}
\caption{Metasurfaces for nonlinear beam deflection. (a-d) Discrete diffraction spectra of TH field in forward direction for four designed metasurfaces operating as beam deflectors: (a) initial set 1, (b) optimum set 2, (c) optimum set 3 of low intensity, and (d) optimum set 3 of high intensity. The intensity is given in arbitrary units. The angle of the first diffraction order is indicated in each spectrum for guidance. The corresponding supercell of the structure and lateral dimensions of the nanodisks used in the simulations are shown in each panel. Orange circles in (a) represent the discrete diffraction spectra of TH field for a metasurface optimized by a gradient descent method in the work of Wang et al. \cite{Wang2018}}
\label{fig:fig5}
\end{figure}

To prove the ability of TH light steering by engineering the $0-2\pi$ phase wavefront using an all-dielectric metasurface, we modeled four metadeflectors using the geometrical parameters designed in the previous sections. All supercells generate a $0-2\pi$ linear phase gradient at the TH wavelength over approximately the same overall supercell length producing a global deflection of the output wave by an angle $\theta$. To estimate the directionality of the TH emission, we calculated discrete diffraction spectra by employing near-field to far-field transformation (NFFFT) for periodic structures\cite{Taflove2005}, as presented in Figure~\ref{fig:fig5}. The corresponding model of the supercell together with the geometry of individual nanodisks and calculated angular value of the first diffraction order are also indicated for each case. As expected, all metasurfaces produce the TH radiation mainly into the positive first diffraction order $+1$ at an angle around $5.6^\circ$ relative to the direction of incident light. These angles differ to some degree owing to slightly distinct lengths of the supercells. Remarkably, the intensities of TH emitted into the 1st diffraction order for our initial supercell 1 and supercell optimized in the work of Wang et al. \cite{Wang2018} have a similar magnitude, as shown in Figure~\ref{fig:fig5}a. Furthermore, the TH efficiencies of all MC-optimized supercells (Figure~\ref{fig:fig5}b-d) are larger than that of the initial supercell 1. In particular, the metadeflectors optimized using the MC procedure in combination with the sampling method shown in Figure~\ref{fig:fig5}b and c demonstrate amplification of the TH intensity by a factor of $\sim 20$ and $\sim 2$, respectively. Moreover, the supercell optimized using MC simulation in conjunction with the manual selection of resonators (Figure \ref{fig:fig5}d) reveals even larger amplification of the TH intensity by a factor of $\sim 500$. Such high enhancement is achieved due to the selection of appropriate resonators for the supercell in a region near a strong optical resonance at the fundamental frequency as can be seen in Figure~\ref{fig:fig4}f (upper-right corner).

Aside from the conversion efficiency of the beam deflectors, a few words should be added about the diffraction efficiency into the 1st diffraction order that varies greatly among the models. The diffraction spectra for the initial and MC-optimized metadeflectors shown in Figure~\ref{fig:fig5}a and b exhibit a high diffraction efficiency in which almost all diffraction orders are well damped except for the +1 order. For the remaining two models shown in Figure~\ref{fig:fig5}c and d, other undesirable diffraction orders notably appear in the diffraction spectra. These directionality diagrams are influenced by an interplay of a number of factors and assumptions present in our approach. First, although our optimization of the TH field generated by the single unit cell comprises coupling between identical neighboring resonators, it ignores the difference in coupling strengths induced by the interaction between neighboring resonators with different dimensions and thus resonances in the supercell. Especially, the coupling strength significantly depends on either the selected resonators operate in an off- or on-resonant state at the fundamental frequency, with the effect being much stronger for the latter. Therefore, only if the neighboring shapes and gaps of nanodisks do not differ much, specifically if the nanodisk shapes and gaps smoothly change within the supercell, the presence of non-identical near-neighbors does not alter the wavefront of the unit cell much. Second, the subwavelength periodicity of the single unit cell at the TH wavelength is essential in order to avoid significant radiation into the higher diffraction orders. Finally, the theoretical maximum of diffraction efficiency is limited by the number of discrete phase steps, in such a manner that it falls as the number of resonators in the supercell decreases \cite{Swanson1991BinaryOT}. 


Indeed, the initial supercell composed of ten nanodisks with the smallest lattice period (\SI{550}{\nano\metre}) and detunded from the resonance peaks (Figure~\ref{fig:fig5}a) reveals the best diffraction efficiency. Interestingly, the supercell shown in Figure~\ref{fig:fig5}b) and consisting of six detuned resonators with a larger periodicity of \SI[mode=text]{916.67}{\nano\metre} also demonstrates a quite high diffraction efficiency, mainly due to the large gaps between the neighboring nanodisks. For both supercells, the excited electric fields inside all nanodisks have similar profiles as shown in Figure S1a,b of Supporting Information, leading to the uniform radiation of the TH field from the individual resonators. On the other hand, the worst accuracy is observed for the supercells composed of six nanodisks with the largest lattice periodicity (\SI{950}{\nano\metre}), small gaps between the resonators, and exhibiting the strong resonant nature at FF (Figure~\ref{fig:fig5}c,d). These factors result in an enhanced near-field interaction with varying coupling strength among the nonidentical nanodisks and, consequently, the strong distortion of the near-field profiles inside the resonators. This leads to undesirable non-uniform radiation of the TH field from the individual nanodisks in the supercell (see Figure S1c,d of Supporting Information).

Nevertheless, the high conversion efficiency of the beam deflectors is our primary goal in this optimization rather than a high level of diffraction efficiency that can be improved inside the presented approach. Overall, these results demonstrate that our combined approach can be successfully used for design and optimization of nonlinear metasurfaces. Although the approach requires a large amount of calculated data, it is simple and can be easily implemented for such high dynamic optimization tasks, especially where a set of optimization parameters, for example the number of resonators used in a final application, is unknown in advance.

\section{Conclusion}
In summary, we have reported a highly efficient and rather simple approach to design all-dielectric metasurfaces for nonlinear wavefront control. By encoding the third harmonic phase and amplitude in a silicon nanodisk and employing MC simulation for amplitude optimization in combination with a sampling method for the selection of appropriate nanodisks, we have demonstrated the full nonlinear $0-2\pi$ phase control with high and truly uniform amplitude. The robustness of our approach has been proved by designing TH beam deflectors in the optical range. We achieved the TH beam deflection into the desired channel with a maximum peak power-independent conversion efficiency of $\eta_{\mathrm{TH}}=6.1 \times 10^{-8}$W$^{-2}$ that is the highest value for silicon metasurfaces reported so far, to the best of our knowledge. The diffraction efficiency of the metadeflectors has been found to depend on the lattice period, gaps between resonators as well as whether they operate off- or on-resonance and can be increased by including additional far-field quantities in the optimization process. We expect that the proposed approach will be widely applied as alternative to more complex optimization algorithms for the design and optimization of metasurfaces for diverse beam-shaping applications.

\section{Methods}
All numerical simulations were performed using COMSOL Multiphysics which employs a finite element method in frequency domain for the calculation of electromagnetic fields \cite{comsol}. Our simulation model for the design and optimization step is reduced to a unit cell with a single resonator from the entire structure of the metasurface (see Figure~\ref{fig:fig1}b). The surface character of the structure is maintained by applying Floquet periodic boundary conditions to the unit cell in the $x$- and $y$-directions to imitate an infinite surface of identical resonators. The boundaries in the $z$-direction are set to open boundary conditions by adding perfectly matched layers (PML) in order to absorb the outgoing electromagnetic waves. Additionally, the model contains lossless integration ports above and below the resonator where the different far-field quantities at the FF and TH are calculated by a near-field to far-field transformation (NFFFT) for periodic structures as \cite{Taflove2005}
\begin{equation}
\mathbf{E}_\text{far-field,m,n}(\omega) = \frac{1}{A_p} \int\limits_0^{x_p}\int\limits_0^{y_p} \mathbf{E}(\omega,x,y,z_0)\exp(j k_{x,m}\,x)\exp(j k_{y,n}\,y)\;\mathrm{d}y\,\mathrm{d}x\\
\label{eq:eq1}
\end{equation}
\noindent where $k_{x,m} = \frac{2\pi m}{x_p}$ and $k_{y,n} = \frac{2\pi n}{y_p}$ are periodicity wave numbers, $A_p=x_p\,y_p$ is the cross-section area of the unit cell. These ports are set to be located in the radiative near-field region of the resonator, in which the evanescent part of the emitted fields is sufficiently small. The nonlinear radiation is calculated via a two steps simulation approach employing an undepleted pump approximation in which the back coupling of the generated TH on the fundamental excitation signal is assumed to be negligible. In the first step, the periodic unit cell with the resonator is excited from the air side with a linear $x$-polarized plane wave propagating in the negative $z$-direction at the fundamental frequency $\omega$. We use a pump wavelength of \SI[mode=text]{1615}{\nano\metre} with input power density of 
\SI{1}{\giga\watt\per\square\centi\meter}. Then, the resulting local electric field and nonlinear polarization induced inside the resonator are calculated. The third-order nonlinear polarization of silicon at frequency $3\omega$ is calculated as $\mathbf{P_{NL}^{(3)}}(3\omega,\mathbf{r}) = \varepsilon_0 \chi^{(3)} (\mathbf{E}(\omega,\mathbf{r})\cdot\mathbf{E}(\omega,\mathbf{r})) \mathbf{E}(\omega,\mathbf{r})$, where the third-order susceptibility tensor $\chi^{(3)}$ is considered as isotropic with a constant value of \SI[per-mode=symbol,mode=text]{2,45e-19}{\square\metre\per\square\volt}. In the second step, the locally generated nonlinear currents inside the resonator $\mathbf{J}_\text{ext}(3\omega,\mathbf{r})=\jmath\,3\omega\mathbf{P_{NL}^{(3)}}(3\omega,\mathbf{r})$ are injected as local current sources in a second simulation run. Subsequently, the generated phase and amplitude of the TH field in the zero-diffracted order ($k_x=k_y=0$) in the transmission direction are calculated using Eq.~\ref{eq:eq1}. The calculation of electromagnetic fields generated by the supercell composed of several resonators are performed in similar way. All calculations use the refractive index of crystalline silicon from the measured dispersion curve provided in Ref.\cite{Green2008} . The refractive index of SiO$_2$ substrate at fundamental and third harmonic frequency is set to $n_s=1.44$ and $1.46$, respectively.

\begin{acknowledgement}
The authors gratefully acknowledge financial support from the Deutsche Forschungsgemeinschaft (DFG) via TRR142 project C05 and computing time support provided by the Paderborn Center for Parallel Computing (PC$^2$).
\end{acknowledgement}

\newpage
\begin{suppinfo}
\subsection{Supplementary note 1: Electric near-fields of supercells for TH beam deflection}
\begin{figure}
        \includegraphics[scale=1]{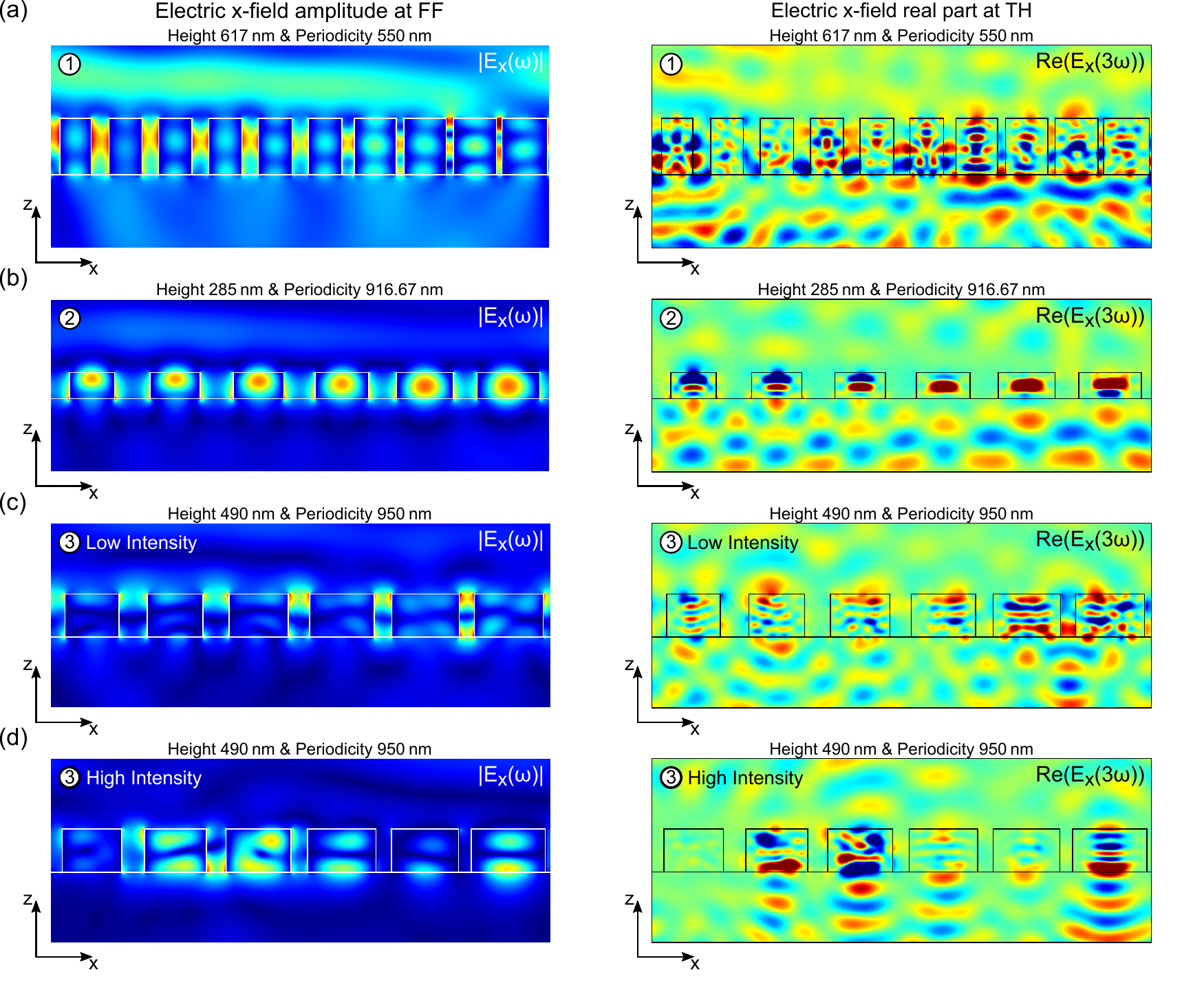}
        \caption{Electric near-field amplitudes $|E_x(\omega)|$ at the fundamental frequency (left) and $\Re[E_x(3\omega)]$ at the third harmonic (right) of the designed supercells for nonlinear beam steering: (a) initial set 1, (b) optimal set 2, (c) optimal set 3 with low intensity, and (d) optimal set 3 with high intensity. The fields are calculated in the $xz$-plane at the center of nanodisks $y=0$.}
        \label{fig:SupplFig5}
\end{figure}
\end{suppinfo}

\bibliography{achemso-demo}
\end{document}